\documentstyle[12pt]{article}

\begin{document}

\begin{center}
\large\bf
Infinite-dimensional Grassmann-Banach algebras
\\[15pt]
\normalsize\bf V.D. Ivashchuk\footnotemark[1]\footnotetext[1]{ivas@rgs.phys.msu.su},
\\[10pt]
\it Center for Gravitation and Fundamental Metrology,
VNIIMS, 3/1 M. Ulyanovoy Str.,
Moscow 117313, Russia  and\\
Institute of Gravitation and Cosmology, PFUR,
Mikhlukho-Maklaya Str. 6, \\ Moscow 117198, Russia

\end{center}

\begin{abstract}

A short review on infinite-dimensional
Grassmann-Banach algebras (IDGBA) is presented.
Starting with the simplest IDGBA  over $K = \mbox{\bf R}$
with $l_1$-norm (suggested by A. Rogers), we define
a more general IDGBA over  complete normed field $K$
with $l_1$-norm and  set of generators  of arbitrary power.
Any $l_1$-type IDGBA may be obtained by action
of Grassmann-Banach functor of projective type
on certain $l_1$-space. In non-Archimedean case
there exists another possibility for constructing
of IDGBA using the Grassmann-Banach functor of injective type.

\end{abstract}

Infinite-dimensional Grassmann-Banach algebras (IDGBA)
and their modifications are key objects for infinite-dimensional versions
of  {\it superanalysis }  (see \cite{DW}-\cite{Khr1} and references
therein). They are generalizations of finite-dimensional
Grassmann algebras to infinite-dimensional Banach case
(for infinite-dimensional topological
Grassmann algebras see also \cite{Ber}).

Any IDGBA is an associative Banach algebra
with unit over some complete normed field $K$ \cite{BS},
whose linear  space $G$ is a Banach space with the
norm $||.||$ satisfying $||a \cdot b || \leq ||a|| ||b||$
for all $a, b \in G$ and $||e|| = 1$, where $e$ is the unit.
(For applications in {\it superanalysis} $K$ should be non-discrete,
i.e. $0 < |v| < 1$ for some $v \in K$, where $|.|$ is
the norm in $K$.) It  contains an infinite subset
of generators $\{ e_{\alpha}, \alpha \in M \} \subset G$,
satisfying
\begin{equation}
\label{0}
e_{\alpha} \cdot e_{\beta} +   e_{\beta} \cdot e_{\alpha} = 0,
\qquad  e_{\alpha}^2 = 0,
\end{equation}
$\alpha, \beta \in M$, where $M$ is some infinite set.
(The second relation in (\ref{1}) follows from the first one
if ${\rm char}K \neq 2$, i.e.  $1_K + 1_K \neq 0_K$.)

The simplest IDGBA  over $K = \mbox{\bf R}$ with $l_1$-norm
was considered by A. Rogers in \cite{R}. In this case $M = \mbox{\bf N}$ and
any element of $a \in G$ can be represented in the form
\begin{equation}
\label{1}
a = a^0 e + \sum_{k \in \mbox{\bf N}} \ \sum_{\alpha_1 < \dots < \alpha_k}
a^{\alpha_1 \dots  \alpha_k} e_{\alpha_1} \cdot \dots \cdot e_{\alpha_k},
\end{equation}
where all $a^0, a^{\alpha_1 \dots  \alpha_k} \in K$  and
\begin{equation}
\label{2}
||a || = |a^0| + \sum_{k \in \mbox{\bf N}} \
\sum_{\alpha_1 < \dots < \alpha_k}
|a^{\alpha_1 \dots  \alpha_k}| < + \infty.
\end{equation}
All series in (\ref{1}) are absolutely convergent w.r.t. the norm
(\ref{2}).

In \cite{I1} a family  of $l_1$-type IDGBA  over a complete normed
field $K$ was suggested. This family
extends  IDGBA from \cite{R} to arbitrary $K$ and arbitrary
infinite number of generators $\{ e_{\alpha}, \alpha \in M \}$.
For linearly ordered set $M$ the relations (\ref{1}) and (\ref{2})
survive, each sum in (\ref{1}) and (\ref{2}) contains
not more than countable number of non-zero terms (AC) (
here and below (AC) means that the axiom of choice \cite{NB} is used).

Here we outline an explicit construction of IDGBA
>from \cite {I1} for arbitrary (not obviously linearly ordered)
index set $M$. Any element of this family
$G(M,K,\langle . \rangle )$ is defined by infinite  set $M$ and
an ordering mapping
$\langle . \rangle: P_0(M) \setminus \{ \emptyset \}
\rightarrow S_0(M)$,
where $P_0(M)$ is the set of all finite subsets of $M$ and
$S_0(M)$  the set of all ordered (non-empty) sets $(s_1, \ldots, s_k)$
of elements from $M$ ($k \in \mbox{\bf N}$). The ordering function
$\langle . \rangle$ obeys the relations
\begin{equation}
\label{3}
\langle \{\alpha_1, \dots,  \alpha_k \} \rangle =
(\alpha_{\sigma(1)}, \dots,  \alpha_{\sigma(k)} ),
\end{equation}
where $\sigma \in S_k$ is some permutation of $\{1, \dots, k \}$,
$k \in \mbox{\bf N}$. The mapping $\langle.\rangle$ does exist (AC).
For linearly ordered $M$ the canonical ordering
function
$\langle . \rangle =\langle . \rangle _0$ is defined by (\ref{3}) with the
inequalities $\alpha_{\sigma(1)} < \dots <  \alpha_{\sigma(k)}$ added.
The vector space of $G(M,K,\langle . \rangle )$ is the Banach space
$G =l_1(P_0(M),K)$  of absolutely summable functions
$a: P_0(M) \rightarrow K$ with the norm
\begin{equation}
\label{4}
||a|| =  \sum_{I \in P_0(M)} |a(I)| < + \infty.
\end{equation}
The operation of multiplication in $G$ is defined
as follows
\begin{equation}
\label{5}
(a \cdot b)(I) =  \sum_{I_1 \cup I_2 = I}
\varepsilon(I_1,I_2) a(I_1) b(I_2),
\end{equation}
$a,b \in G$, $I \in P_0(M)$, where
$\varepsilon: P_0(M) \times P_0(M) \rightarrow K$ is $\varepsilon$-symbol:
\begin{eqnarray}
\label{6}
\varepsilon(I_1,I_2) = &0_K, \ &{\rm if} \ I_1 \cap I_2
                              \neq \emptyset,  \\
                       &\ 1_K, \ &{\rm if} \ I_1 = \emptyset,
                               \  {\rm or} \ I_2 = \emptyset, \nonumber \\
                       &\varepsilon_{\sigma}, \ &{\rm otherwise} \nonumber,
\end{eqnarray}
where $\varepsilon_{\sigma} = \pm 1_K$ is the parity of the
permutation $\sigma$:
$(\langle I_1 \rangle,\langle I_2 \rangle ) \mapsto
\langle I_1 \cup I_2 \rangle$.  For any $a \in G$ we get $a = \sum_{I \in
P_0(M)} a(I)e_I$, where $(e_I, I \in P_0(M))$ is the Shauder basis in $G$
defined by the relations: $e_I(J) = \delta_I^J$ for $I,J \in P_0(M)$. The
unit is $e = e_{\emptyset}$ and generators are $e_{\alpha} = e_{\{\alpha
\}}$, $\alpha \in M$.  Decomposition (\ref{1}) is valid for general
ordering function $\langle.\rangle$ if $a^0 = a({\emptyset})$,
$a^{\alpha_1 \dots  \alpha_k} = a(\{\alpha_1, \dots,  \alpha_k \})$
and relations $\alpha_1 < \dots < \alpha_k$ are understood as
$(\alpha_1, \dots, \alpha_k) \in
\langle P_0(M) \rangle $ ($\langle P_0(M) \rangle$ is the image
of $P_0(M)$ under the mapping $\langle . \rangle $).

Banach algebra (BA) $G(M,K,\langle . \rangle )$ depends essentially only
upon the cardinal number $[M]$ of the set $M$, i.e.
$G(M_1,K,\langle . \rangle _1)$ and $G(M_2,K,\langle . \rangle _2)$ are
isomorphic (in the category of BA) if and only if $[M_1] = [M_2]$ (AC)
\cite{I1}.  The  Banach space of  $G(M,K,\langle . \rangle )$ may be
decomposed into a sum of two closed subspaces
\begin{equation}
\label{7}
G = G_0 \oplus G_1,
\end{equation}
where
$G_i = \{a \in G| a(I) = 0_K, I \in  P_0(M), |I| \equiv i+1
({\rm mod} \ 2 ) \}$, $i = 0,1$.
(The subspace $G_0$ ($G_1$) consists of sums of even (odd) monoms  in
(\ref{1})).  BA $G(M,K,\langle . \rangle )$ with the decomposition
(\ref{7}) is a supercommutative (Banach) superalgebra
\begin{eqnarray}
\label{8a}
a \cdot b = (-1_K)^{ij} b \cdot a, \quad a \in G_i, \ b \in G_j, \\
\label{8b}
G_i \cdot G_j \subset G_{i+j} ({\rm mod} \ 2),
\end{eqnarray}
$i,j =0,1$. The odd subspace $G_1$ has
trivial  (right) annihilator \cite{I1}
\begin{equation}
\label{9}
{\rm Ann}(G_1) \equiv \{a \in G|  G_1 \cdot a = \{0 \} \}
= \{0 \}.
\end{equation}
This relation is an important one for applications
in {\it superanalysis}, since it provides  the
definitions of all superderivatives as elements of $G$.
Note that  any non-trivial (non-zero)
associative supercommutative superalgebra over $K$,
${\rm char}K \neq 2$, is infinite-dimensional \cite{I1}
(for $K = \mbox{\bf R}, \mbox{\bf C}$ see
also \cite{Khr1}).

Another important (e.g. for applications in {\it superanalysis})
proposition is the following one \cite{R,I2}: in $G(M,K,\langle.\rangle)$
the element $a$ is invertible if and only if
$a^0 = a(\emptyset) \neq 0_K$.
(In \cite{I2} an explicit expression for inverse element $a^{-1}$ was
obtained.)

IDGBA with $l_1$-norm forms a special subclass of
more general family of IDGBA over $K$ \cite{I3}, namely,
\begin{equation}
\label{9a}
G(M,K,\langle . \rangle ) \cong \hat{{\cal G}}(l_1(M,K)),
\end{equation}
where $\hat{{\cal G}} = \hat{{\cal G}}_K$ is the Grassmann-Banach
{\it functor} of projective type \cite{I3}.
Here
\begin{equation}
\label{10}
\hat{{\cal G}}(E) = \hat{{\cal T}}(E)/\hat{I},
\end{equation}
where $\hat{{\cal T}}(E)$ is a tensor BA
of projective type  corresponding to infinite-dimensional
projectively proper Banach space $E$ over $K$
and $\hat{I}$ is a closed ideal generated by
the subset $\{ a^2, a \in E \}$.
Banach space $E$ over $K$ is called projectively proper
if all projective seminorms $p_k: E^{\otimes k} =
E \otimes \dots  \otimes E \ ( k-{\rm times}) \rightarrow \mbox{\bf R}$,
$k \geq
2$, are norms \cite{I3}. For $K = \mbox{\bf R}, \mbox{\bf C}$ any $E$ is
projectively proper \cite{Khel}.  Tensor Banach functor $\hat{{\cal T}}
=\hat{{\cal T}}_K$ was defined in \cite{I3} (for  tensor BA without unit
over $K = \mbox{\bf C}$ see \cite{Yak}).
The Banach space of $\hat{{\cal
T}}(E)$ is a $l_1$-sum of projective tensor powers of $E$
\begin{equation}
\label{11}
\hat{T}(E) =
\hat{\oplus} \sum_{i = 0}^{\infty} \hat{T}_i(E),
\end{equation}
where $\hat{T}_0(E) = K$, $\hat{T}_1(E) =E$ and $\hat{T}_k(E) =
E \hat{\otimes} \dots  \hat{\otimes}E$ ($k$-times)
are projective tensor products,
$k \geq 2$. The norm of $a = (a_0,a_1, \ldots) \in
\hat{T}(E)$ is $||a || = ||a_0 ||_0 + ||a_1 ||_1 + \ldots$,
where $a_i \in
\hat{T}_i(E)$ and  $||.||_i$ is  projective norm in $\hat{T}_i(E)$, $i
= 0,1, \ldots$.

For non-Archimedean field $K$ satisfying: $|x+y| \leq {\max}(|x|,|y|)$,
$x,y \in K$, there exists another possibility for constructing of IDGBA
\cite{I4}. The Grassmann-Banach  functor of injective type
$\check{{\cal G}} = \check{{\cal G}}_K$ is defined
for certain subclass of injectively proper
non-Archimedean Banach spaces over $K$.
Banach space $E$ over $K$ is called injectively proper
if the injective seminorms $w_k: E^{\otimes k} \rightarrow \mbox{\bf R}$,
$k \geq
2$, are norms \cite{I4}. In this case (\ref{10}) is modified as follows
\begin{equation}
\label{10na}
\check{{\cal G}}(E) = \check{{\cal T}}(E)/\check{I},
\end{equation}
where $\check{{\cal T}}(E)$ is tensor BA
of injective type  corresponding to $E$
and $\check{I}$ is a closed ideal generated by
the subset $\{ a^2, a \in E \}$.
The Banach space of $\check{{\cal
T}}(E)$ is a $l_{\infty}$-sum of injective tensor powers of $E$
\begin{equation}
\label{11n}
\check{T}(E) =
\check{\oplus} \sum_{i = 0}^{\infty} \check{T}_i(E),
\end{equation}
where $\check{T}_0(E) = K$, $\check{T}_1(E) =E$ and $\check{T}_k(E) =
E \check{\otimes} \dots  \check{\otimes}E$ ($k$-times)
are injective tensor products,
$k \geq 2$. The norm of $a = (a_0,a_1,
\ldots) \in \check{T}(E)$ is $||a || = {\rm sup} (||a_0 ||_0,||a_1 ||_1,
\ldots)$,
where $a_i \in \check{T}_i(E)$ and $||.||_i$ is injective norm in
$\check{T}_i(E)$, $i = 0,1, \ldots$.
For $l_{\infty}$-spaces we have an isomorphism of BA
\begin{equation}
\label{9an}
G_{\infty}(M,K,\langle . \rangle ) \cong \check{{\cal G}}(l_{\infty}(M,K)),
\end{equation}
where
$G_{\infty}(M,K,\langle . \rangle )$ is the Grassmann-Banach algebra
with the Banach space  $l_{\infty}(P_0(M),K)$
and the multiplication defined in (\ref{5}).
Here $l_{\infty}(P_0(M),K)$  is the Banach space of bounded
functions $a: P_0(M) \rightarrow K$ with the norm
\begin{equation}
\label{4na}
||a||_{\infty} =  {\rm sup}(|a(I)|, I \in P_0(M)) .
\end{equation}

For applications in superanalysis the following supercommutative
Banach superalgebras may be also used : $B \hat{\otimes} {\cal G}$.
Here $B$ is an associative
commutative BA with unit over $K$, and  ${\cal G}$ is IDGBA. For
${\cal G} =G(M,K,\langle . \rangle )$
we have an isomorphism of BA : $B \hat{\otimes} G(M,K,\langle . \rangle )
\cong  G(M,B,\langle.\rangle)$, where $G(M,B,\langle . \rangle )$  is
obtained from $G(M,K, \langle . \rangle )$ by the replacement $K \mapsto
B$ (for $M = {\bf N}$ see also \cite{Khr}).

For non-Archimedean $B$, ${\cal G}$ and $K$
another Banach superalgebra may be also considered : $B \check{\otimes}
{\cal G}$. In this case
$B \check{\otimes} G_{\infty}(M,K,\langle . \rangle ) \cong
G_{\infty}(M,B,\langle.\rangle)$,
where $G_{\infty}(M,B,\langle . \rangle )$  is obtained from
$G_{\infty}(M,K, \langle .\rangle )$ by the replacement $K \mapsto B$.

\end{document}